\documentstyle[11pt]{article}

\begin{document}

\setcounter{page}{1}

\pagestyle{plain} \vspace{1cm}

\begin{center}
\Large{\bf Non-Minimal Braneworld Inflation after \emph{Planck} }\\
\small \vspace{1cm} {\bf Kourosh
Nozari\footnote{knozari@umz.ac.ir}}\quad and\quad {\bf Narges Rashidi\footnote{n.rashidi@umz.ac.ir}}\\
\vspace{0.5cm} Department of Physics, Faculty of Basic Sciences,
University of Mazandaran,\\
P. O. Box 47416-95447, Babolsar, IRAN
\end{center}

\begin{abstract}
The recently released Planck data have constrained 4-dimensional
inflationary parameters even more accurately than ever. We consider
an extension of the braneworld model with induced gravity and a
non-minimally coupled scalar field on the brane. We constraint the
inflation parameters in this setup, by adopting six types of
potential, in confrontation with the joint Planck+WMAP9+BAO data. We
show that a potential of the type
$V(\varphi)=V_{0}\exp(-\beta\varphi)$ has the best fit with newly
released observational data.
\end{abstract}
\newpage

\section{Introduction}

The early time inflationary stage of the universe evolution can
address successfully some problems of the standard big bang
cosmology such as the flatness, horizon and relics problems. Several
theoretical approaches have been proposed to model an inflationary
universe. One of the simple inflationary models is the one in which
the universe is filled with a canonical scalar field (Inflaton)
[1-8]. In order to run inflation successfully, the potential energy
of the inflaton should dominate the kinetic energy of the field.
But, this simple inflation paradigm, suffers by itself from several
problems with no concrete solutions [6,9]. So, other inflationary
models such as the braneworld models [10-14], models with
non-minimally coupled inflaton field [15-22], modified gravity
[23-25] and so on, with a wide range of potentials and with
successes and shortcomings, have attracted much attention these
years. In this regard, a large number of models have been proposed
but this does not mean that all of these models are observationally
viable. For a model to be viable, its consistency with observational
data is inevitable. Also a successful inflationary model is the one
that can provide a mechanism for generating the initial fluctuations
and perturbations in the early universe and therefore seed the
formation of the structures in the universe. In this regard,
fluctuation in the scalar field leads to the scalar power spectra
and fluctuation in the transverse and traceless parts of the metric
leads to the tensor power spectra [1-8]. The scalar power spectrum
of the perturbation is nearly scale-invariant, that is, the spectral
index is about unity. The exact value of the spectral index can be
obtained by using the observational data. The running of the
spectral index and also, the ratio between the amplitudes of tensor
and scalar perturbations (tensor-to-scalar ratio) are others
inflationary parameters which can be constrained observationally.
The values of these parameters calculated in a specific inflation
model and compared with observational data, are powerful probes to
save or rule out the inflation model under consideration. The
recently released Planck observational data, have constrained the
inflationary parameters even more accurately than previous probes
[26]. Those models which their inflationary parameters lie within
the $68\%$ or $95\%$ CL with Planck data can be considered as
observationally viable proposals for inflation. In Ref. [26], the
authors have considered some types of the potential and have found
some constraints on the models parameters. For example, they have
shown that a 4-dimension model with a linear potential lies well
inside the Planck data while a 4-dimension model with an exponential
potential is well outside the Planck data. Also $R^{2}$ inflation
based on modified gravity has a very good agreement with Planck
data.

In the present paper, following our recent work [14], we consider a
branworld model with a non-minimal coupling between the induced
Ricci scalar and the inflaton field on the brane. By adopting six
types of potential, we perform a numerical analysis on the
inflationary parameters of this model. In the background of the
Planck+WMAP9+BAO data, the parameters of the model can be
constrained tightly. We find that some potentials that are suitable
in 4-dimensional case in comparison with observational data, cannot
lead to a successful inflationary model in 5-dimension in the
minimal case. However, by considering a non-minimal coupling between
the scalar field and induced gravity term on the brane, we achieve
good results in some ranges of the non-minimal coupling parameter.
On the other hand, some potentials which are not compatible with
Planck data in a 4-dimensional model, lead to good results in a
branworld setup.\\

\section{The setup}

In this section we present preliminaries and the mathematical
framework of our setup based on our recent work [14]. We consider a
warped DGP model in which an inflaton field is non-minimally coupled
with the induced Ricci scalar on the brane and its action is given
by the following expression
\begin{equation}
S=\frac{1}{2\kappa_{5}^{2}}\int d^{5}x\sqrt{-g^{(5)}}\bigg[R^{(5)}-
2\Lambda_{5}\bigg]+\int_{brane} d^{4} x
\sqrt{-q}\bigg[\frac{1}{2\kappa_{4}^{2}}R
+\frac{f(\varphi)}{2}\,R-\lambda-
\frac{1}{2}q^{\mu\nu}\partial_{\mu}\varphi\partial_{\nu}\varphi-V(\varphi)\bigg]
 \label{1}
\end{equation}
where $\kappa_{5}^{2}$\,,$R$\,,$R^{(5)}$\,,$\lambda$\,and
$\Lambda_{5}$ are the five dimensional gravitational constant, the
induced Ricci scalar on the brane, 5-dimensional Ricci scalar, the
brane tension and the bulk cosmological constant respectively. $q$
is the trace of the brane metric, $q_{\mu\nu}$. Also, $f(\varphi)$
shows the non-minimal coupling of the scalar field with induced
gravity on the brane. The action (\ref{1}) in a FRW background gives
the generalized cosmological dynamics in this setup by the following
Friedmann equation
\begin{equation}
H^{2}=\frac{\kappa_{4}^{2}}{3}\rho_{\varphi}+\frac{\kappa_{4}^{2}}{3}\lambda
+\frac{2\kappa_{4}^{4}}{\kappa_{5}^{4}}
\pm\frac{2\kappa_{4}^{2}}{\kappa_{5}^{2}}\,\sqrt{\frac{\kappa_{4}^{4}}{\kappa_{5}^{4}}+
\frac{\kappa_{4}^{2}}{3}\rho_{\varphi}+
\frac{\kappa_{4}^{2}}{3}\lambda-\frac{\Lambda_{5}}{6}-\frac{\mathcal{C}}{a^{4}}}.\hspace{0.5cm}
\label{2}
\end{equation}
where $\rho_{\varphi}$ is the energy-density corresponding to the
non-minimally coupled scalar field given by
\begin{equation}
\rho_{\varphi}=\frac{1}{2}\dot{\varphi}^{2}+V(\varphi)-6f'(\varphi)H\dot{\varphi}\,.
\label{3}
\end{equation}
Also, the corresponding pressure is defined as follows
\begin{equation}
p_{\varphi}=\frac{1}{2}\dot{\varphi}^{2}-V(\varphi)
+2f'(\varphi)\ddot{\varphi}+4f'(\varphi)H\dot{\varphi}
+2f''(\varphi)\dot{\varphi}^{2}\,, \label{4}
\end{equation}
where a prime represents the derivative with respect to the scalar
field and a dot marks derivative with respect to the cosmic time.

Here we are interested in the case where inflationary dynamics is
driven by a scalar field with a self-interacting potential. So, the
effective cosmological constant on the brane which is defined as
\begin{equation}
\Lambda_{eff}=\kappa_{4}^{2}\lambda+\frac{6\kappa_{4}^{4}}{\kappa_{5}^{4}}
\pm\frac{\sqrt{6}\kappa_{4}^{4}}{\kappa_{5}^{4}}
\sqrt{\Big(2\kappa_{4}^{2}\lambda-\Lambda_{5}\Big)\frac{\kappa_{5}^{4}}{\kappa_{4}^{4}}+6}\,\label{5}
\end{equation}
should be equal to zero, so that
\begin{equation}
\Lambda_{5}=-\frac{\kappa_{5}^{4}}{6}\lambda^{2}\,.\label{6}
\end{equation}
Now, the Friedmann equation (\ref{2}) can be rewritten as follows
\begin{equation}
H^{2}=\frac{\kappa_{4}^{2}}{3}\rho_{\varphi}+\frac{\kappa_{4}^{2}}{3}\lambda
+\frac{2\kappa_{4}^{4}}{\kappa_{5}^{4}}
\pm\frac{2\kappa_{4}^{2}}{\kappa_{5}^{2}}\,\sqrt{\frac{\kappa_{4}^{4}}{\kappa_{5}^{4}}+
\frac{\kappa_{4}^{2}}{3}\rho_{\varphi}+
\frac{\kappa_{4}^{2}}{3}\lambda-\frac{\kappa_{5}^{4}}{36}\lambda^{2}-\frac{\mathcal{C}}{a^{4}}}\,.\label{7}
\end{equation}
In our setup, the slow-roll parameters ($ \epsilon\equiv
-\frac{\dot{H}}{H^{2}}$ and $\eta\equiv
-\frac{1}{H}\frac{\ddot{H}}{\dot{H}}$), in the slow-roll
approximation ($\dot{\varphi}^{2}\ll V(\varphi)$ and
$\ddot{\varphi}\ll3H\dot{\varphi}$ ), take the following form
respectively
\begin{equation}
\epsilon \simeq \frac{1}{2\kappa_{4}^{2}}\frac{V'^2}{V^2}\times
{\cal{A}}(\varphi)\,,\label{8}
\end{equation}
and
\begin{equation}
\eta \simeq \frac{1}{\kappa_{4}^{2}}\frac{V''}{V}\times
{\cal{B}}(\varphi)\,,\label{9}
\end{equation}
where by definition
\begin{eqnarray}
{\cal{A}}(\varphi)=\Bigg(\frac{1}{V'}
-\frac{f'R}{2V'^{2}}\Bigg)\Bigg(V'-2f'f''R+2f''V'+2f'V''\Bigg)\hspace{5.5cm}\nonumber\\
\times\frac{1\pm\frac{\kappa_{4}^{2}}{\kappa_{5}^{2}}
\frac{1-\frac{\mathcal{C}}{a^{4}}\frac{36H^{2}}{\kappa_{4}^{2}\Big(V'
-\frac{f'R}{2}\Big)\Big(V'-2f'f''R+2f''V'+2f'V''\Big)}}{\sqrt{\frac{\kappa_{4}^{4}}{\kappa_{5}^{4}}+
\frac{\kappa_{4}^{2}}{3}V-\frac{\kappa_{4}^{2}}{3}f'^{2}R+\frac{2\kappa_{4}^{2}}{3}f'V'+
\frac{\kappa_{4}^{2}}{3}\lambda-\frac{\kappa_{5}^{4}}{36}\lambda^{2}
-\frac{\mathcal{C}}{\hat{a}^{4}}}}}{\Bigg[1
+\frac{\lambda}{V}-\frac{f'^{2}R}{V}+\frac{2f'V'}{V}+\frac{6\kappa_{4}^{2}}{\kappa_{5}^{4}V}
\pm\frac{6}{\kappa_{5}^{2}V}\,\sqrt{\frac{\kappa_{4}^{4}}{\kappa_{5}^{4}}+
\frac{\kappa_{4}^{2}}{3}V-\frac{\kappa_{4}^{2}}{3}f'^{2}R+\frac{2\kappa_{4}^{2}}{3}f'V'+
\frac{\kappa_{4}^{2}}{3}\lambda-\frac{\kappa_{5}^{4}}{36}\lambda^{2}
-\frac{\mathcal{C}}{a^{4}}}\,\,\Bigg]^2},\label{10}
\end{eqnarray}
and
\begin{eqnarray}
{\cal{B}}(\varphi)=\Bigg(1
-\frac{f''R}{2V''}\Bigg)\hspace{12cm}\nonumber\\
\times\Bigg\{\frac{1}{1
+\frac{\lambda}{V}-\frac{f'^{2}R}{V}+\frac{2f'V'}{V}+\frac{6\kappa_{4}^{2}}{\kappa_{5}^{4}V}
\pm\frac{6}{\kappa_{5}^{2}V}\,\sqrt{\frac{\kappa_{4}^{4}}{\kappa_{5}^{4}}+
\frac{\kappa_{4}^{2}}{3}V-\frac{\kappa_{4}^{2}}{3}f'^{2}R+\frac{2\kappa_{4}^{2}}{3}f'V'+
\frac{\kappa_{4}^{2}}{3}\lambda-\frac{\kappa_{5}^{4}}{36}\lambda^{2}
-\frac{\mathcal{C}}{a^{4}}}}\Bigg\}\,.\label{11}
\end{eqnarray}
Equations (\ref{10}) and (\ref{11}) reflect the non-minimal coupling
of the scalar field and induced gravity on the brane and also,
braneworld nature of the setup.

The number of e-folds form initial of inflation ($t_{i}$) until the
time where the inflation ends ($t_{f}$), is given by
$N=\int_{t_{i}}^{t_{f}} H dt$. For a warped DGP model with a
non-minimally coupled scalar field on the brane, the number of
e-folds takes the following expression
\begin{eqnarray}
N=\int_{\varphi_{hc}}^{\varphi_{f}}
\Bigg(\frac{3V}{V'}\Bigg)\Bigg(\frac{V'}{\frac{1}{2}f'R-V'}\Bigg)\Bigg[\frac{\kappa_{4}^{2}}{3}
+\frac{1}{V}\bigg(\frac{\kappa_{4}^{2}}{3}\lambda-\frac{\kappa_{4}^{2}f'^{2}R}{3}+\frac{2\kappa_{4}^{2}f'V'}{3}
+\frac{2\kappa_{4}^{4}}{\kappa_{5}^{4}}\hspace{3cm}\nonumber\\
\pm\frac{2\kappa_{4}^{2}}{\kappa_{5}^{2}}\,\sqrt{\frac{\kappa_{4}^{4}}{\kappa_{5}^{4}}+
\frac{\kappa_{4}^{2}}{3}V(\varphi)-\frac{\kappa_{4}^{2}}{3}f'^{2}R+\frac{2\kappa_{4}^{2}}{3}f'V'+
\frac{\kappa_{4}^{2}}{3}\lambda-\frac{\kappa_{5}^{4}}{36}\lambda^{2}
-\frac{\mathcal{C}}{a^{4}}}\,\bigg)\Bigg]d\varphi\,\,.\label{12}
\end{eqnarray}
Note that, in equation (\ref{12}), $\varphi_{hc}$ marks the value of
$\varphi$ when the universe scale crosses the Hubble horizon during
inflation and $\varphi_{f}$ denotes the value of $\varphi$ when the
universe exits the inflationary phase.

The spectrum of perturbations produced due to quantum fluctuations
of the fields about their homogeneous background values is the
important way to test the viability of inflationary models. With the
following perturbed FRW metric
\begin{equation}
ds^{2}=-\big(1+2\Phi\big)dt^{2}+a^{2}(t)\big(1-2\Psi\big)\delta_{i\,j}\,dx^{i}dx^{j},\label{13}
\end{equation}
which is defined in a longitudinal gauge [27-29], the scalar
spectral index, which describes the scale-dependence of the
perturbations, is given by the following expression
\begin{equation}
n_{s}-1=\frac{d \ln A_{S}^{2}}{d \ln k}\,,\label{14}
\end{equation}
where $d \ln k(\varphi)=d N(\varphi)$. When $n_{s}$ is unity, the
power spectrum of the perturbation is scale invariant. In our warped
DGP setup, we obtain the scalar spectral index within the slow-roll
approximations as follows
\begin{eqnarray}
n_{s}=1-3\epsilon+\frac{2}{3}\eta\hspace{12cm}\nonumber\\
+\Bigg[\frac{-2\Big(V'-\frac{1}{2}f'R\Big)
\Big(\frac{\kappa_{5}^{2}}{6}\rho_{\varphi}\Big)}{6\Big(1+\frac{\kappa_{5}^{2}}{18}\rho_{\varphi}f\Big)
\Big(\rho_{\varphi}+\lambda+\frac{6\kappa_{4}^{2}}{\kappa_{5}^{4}}
\pm\frac{6}{\kappa_{5}^{2}}\,\sqrt{\frac{\kappa_{4}^{4}}{\kappa_{5}^{4}}+
\frac{\kappa_{4}^{2}}{3}\rho_{\varphi}+
\frac{\kappa_{4}^{2}}{3}\lambda-\frac{\kappa_{5}^{4}}{36}\lambda^{2}-\frac{\mathcal{C}}{a^{4}}}
\Big)}\frac{9H^{2}}{\frac{1}{2}f'R-V'}+\frac{2V''}{V'}\Bigg]\hspace{0cm}\nonumber\\
\times\Bigg[\frac{\kappa_{4}^{2}}{3}
+\frac{1}{V}\bigg(\frac{\kappa_{4}^{2}}{3}\lambda-\frac{\kappa_{4}^{2}f'^{2}R}{3}+\frac{2\kappa_{4}^{2}f'V'}{3}
+\frac{2\kappa_{4}^{4}}{\kappa_{5}^{4}}\hspace{4cm}\nonumber\\
\pm\frac{2\kappa_{4}^{2}}{\kappa_{5}^{2}}\,\sqrt{\frac{\kappa_{4}^{4}}{\kappa_{5}^{4}}+
\frac{\kappa_{4}^{2}}{3}V(\varphi)-\frac{\kappa_{4}^{2}}{3}f'^{2}R+\frac{2\kappa_{4}^{2}}{3}f'V'+
\frac{\kappa_{4}^{2}}{3}\lambda-\frac{\kappa_{5}^{4}}{36}\lambda^{2}-\frac{\mathcal{C}}{a^{4}}}\,\bigg)\Bigg]^{-1}
\Bigg(\frac{V'-\frac{1}{2}f'R}{3V}\Bigg).\label{15}
\end{eqnarray}
The running of the spectral index in our setup is given by
\begin{eqnarray}
\alpha=\frac{d n_{s}}{d \ln k}
=6\epsilon^{2}+2\epsilon\eta-\left[\frac{\frac{1}{2}f''R-V''}{H^{4}}\right]\left[V'''-\frac{1}{2}f'''R\right]
+\frac{\frac{1}{2}\big(f''R-2V''\big)^{2}}{\big(\frac{1}{2}f'R-V'\big)^{2}}\hspace{3cm}\nonumber\\
+\frac{f'''R-2V'''}{\frac{1}{2}f'R-V'}-\frac{3\ddot{H}}{H^{2}}+\Bigg[\dot{H}+\frac{V''}{V'}\bigg(\Big(V''
-\frac{1}{2}f''R\Big)\Big(1+\frac{3H^{4}V'}{2V''}\Big)+\dot{H}
\bigg)\Bigg]\hspace{3cm}\nonumber\\
\times\Bigg[\frac{-4\Big(V'-\frac{1}{2}f'R\Big)
\Big(\frac{\kappa_{5}^{2}}{6}\rho_{\varphi}\Big)}{6\Big(1+\frac{\kappa_{5}^{2}}{18}\rho_{\varphi}f\Big)
\Big(\rho_{\varphi}+\lambda+\frac{6\kappa_{4}^{2}}{\kappa_{5}^{4}}
\pm\frac{6}{\kappa_{5}^{2}}\,\sqrt{\frac{\kappa_{4}^{4}}{\kappa_{5}^{4}}+
\frac{\kappa_{4}^{2}}{3}\rho_{\varphi}+
\frac{\kappa_{4}^{2}}{3}\lambda-\frac{\kappa_{5}^{4}}{36}\lambda^{2}-\frac{\mathcal{C}}{a^{4}}}
\Big)}\Bigg]\left[\frac{V'-\frac{1}{2}f'R}{3H^{4}}\right]+{\cal{G}}'\label{16}
\end{eqnarray}
The tensor perturbations amplitude of a given mode, in the time of
Hubble crossing, are given by
\begin{equation}
A_{T}^{2}=\frac{4\kappa_{4}^{2}}{25\pi}H^{2}\Bigg|_{k=aH}\,.\label{17}
\end{equation}
In our setup and within the slow-roll approximation, we find
\begin{eqnarray}
A_{T}^{2}=\frac{4\kappa_{4}^{2}}{25\pi}V\Bigg[\frac{\kappa_{4}^{2}}{3}
+\frac{1}{V}\bigg(\frac{\kappa_{4}^{2}}{3}\lambda-\frac{\kappa_{4}^{2}f'^{2}R}{3}+\frac{2\kappa_{4}^{2}f'V'}{3}
+\frac{2\kappa_{4}^{4}}{\kappa_{5}^{4}}\hspace{6cm}\nonumber\\
\hspace{4cm}\pm\frac{2\kappa_{4}^{2}}{\kappa_{5}^{2}}\,\sqrt{\frac{\kappa_{4}^{4}}{\kappa_{5}^{4}}+
\frac{\kappa_{4}^{2}}{3}V(\varphi)-\frac{\kappa_{4}^{2}}{3}f'^{2}R+\frac{2\kappa_{4}^{2}}{3}f'V'+
\frac{\kappa_{4}^{2}}{3}\lambda-\frac{\kappa_{5}^{4}}{36}\lambda^{2}}\,\bigg)\Bigg]\,.\hspace{1cm}\label{18}
\end{eqnarray}
The tensor(gravitational wave) spectral index which is given by
\begin{equation}
n_{T}=\frac{d \ln A_{T}^{2}}{d \ln k}\,,\label{19}
\end{equation}
in our model and in terms of the slow-roll parameters, can be
expressed as
\begin{equation}
n_{T}=-2\epsilon\,.\label{20}
\end{equation}

Another important inflationary parameter which can be compared with
observation is the ratio between the amplitudes of tensor and scalar
perturbations (tensor-to-scalar ratio). This ratio is given by
\begin{equation}
r\equiv\frac{A_{T}^{2}}{A_{S}^{2}}\simeq\frac{8\pi\kappa_{4}^{2}}{25}\frac{\exp
\Bigg(\int - {\cal{G}}d\varphi\Bigg)}{{\cal{C}}^{2}
V'^{2}k^{3}}\,,\label{21}
\end{equation}
where
\begin{equation}
{\cal{G}}=\frac{-2\Big(V'-\frac{1}{2}f'R\Big)
\Big(\frac{\kappa_{5}^{2}}{6}\rho_{\varphi}+\frac{1}{\dot{\varphi}\delta\varphi}\int\delta
E_{i}^{0}\,dx^{i}\Big)}{6\Big(1+\frac{\kappa_{5}^{2}}{18}\rho_{\varphi}f\Big)
\Big(\rho_{\varphi}+\lambda+\frac{6\kappa_{4}^{2}}{\kappa_{5}^{4}}
\pm\frac{6}{\kappa_{5}^{2}}\,\sqrt{\frac{\kappa_{4}^{4}}{\kappa_{5}^{4}}+
\frac{\kappa_{4}^{2}}{3}\rho_{\varphi}+
\frac{\kappa_{4}^{2}}{3}\lambda-\frac{\kappa_{5}^{4}}{36}\lambda^{2}-\frac{\mathcal{C}}{a^{4}}}
\Big)}-\frac{f''R-2V''}{\frac{1}{2}f'R-V'}+\frac{2V''}{V'}.\label{22}
\end{equation}

After presenting the cosmological dynamics equations, we perform
numerical analysis on the inflationary parameters of the warped DGP
setup with a non-minimally coupled scalar field on the brane. To
this end, we adopt the non-minimal coupling function as
$f(\varphi)=\xi\varphi^{2}$, where $\xi$ is a constant. We consider
some types of potential, substitute them in the integral of equation
(\ref{12}) and solve this equation. After that we find
$\varphi_{hc}$ in terms of $N$. We substitute $\varphi_{hc}$ in
$n_{s}$, $r$ and $\alpha$ and then by varying $\xi$ we find, for
each given values of $N$, the range of $\xi$ where these
inflationary parameters are compatible with Planck+WMAP9+BAO data.

\section{Observational constraint}

\subsection{$V(\varphi)=\frac{\sigma}{2}\varphi^{2}$}
\begin{figure}[htp]
\begin{center}\includegraphics{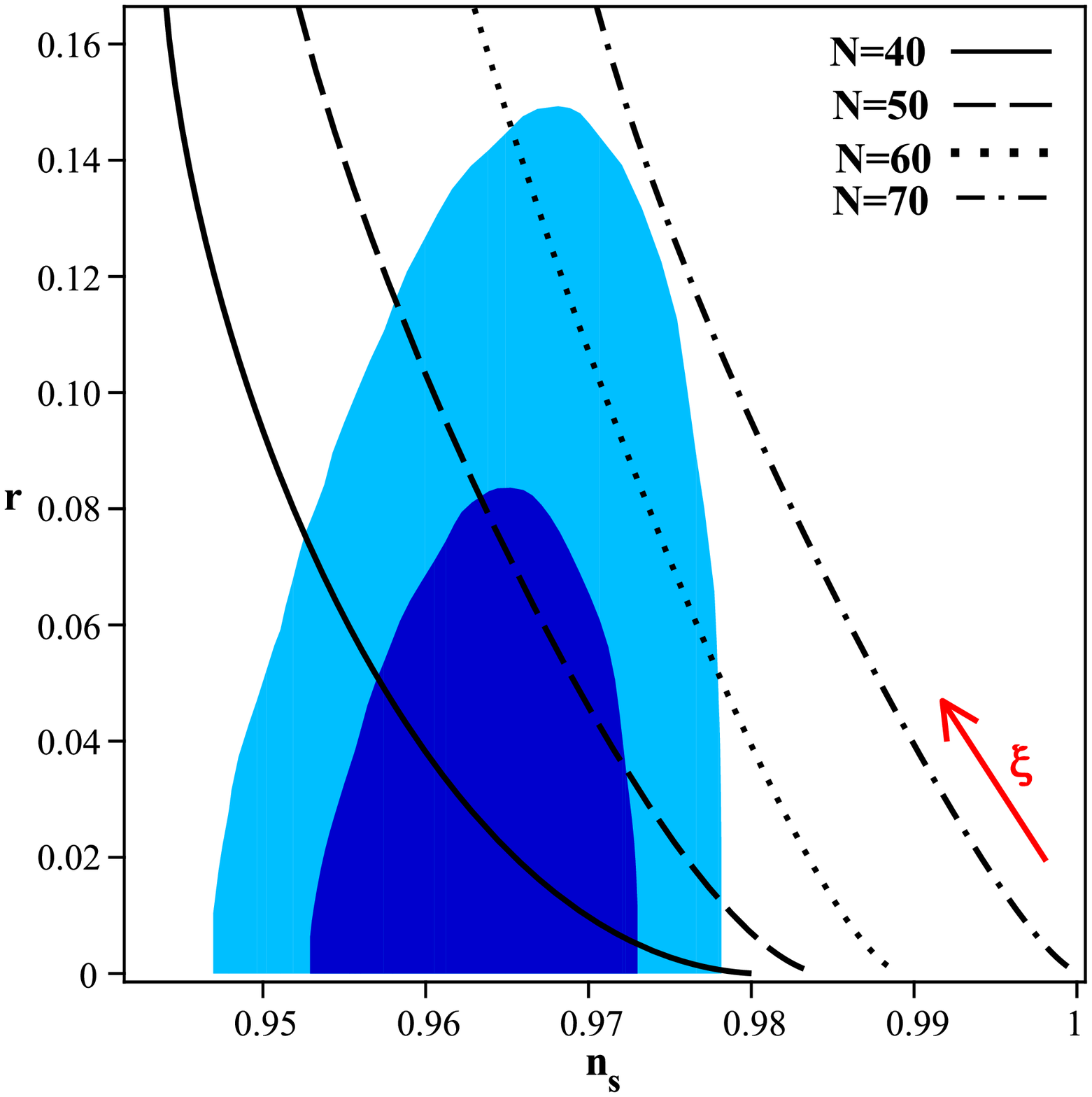} \includegraphics{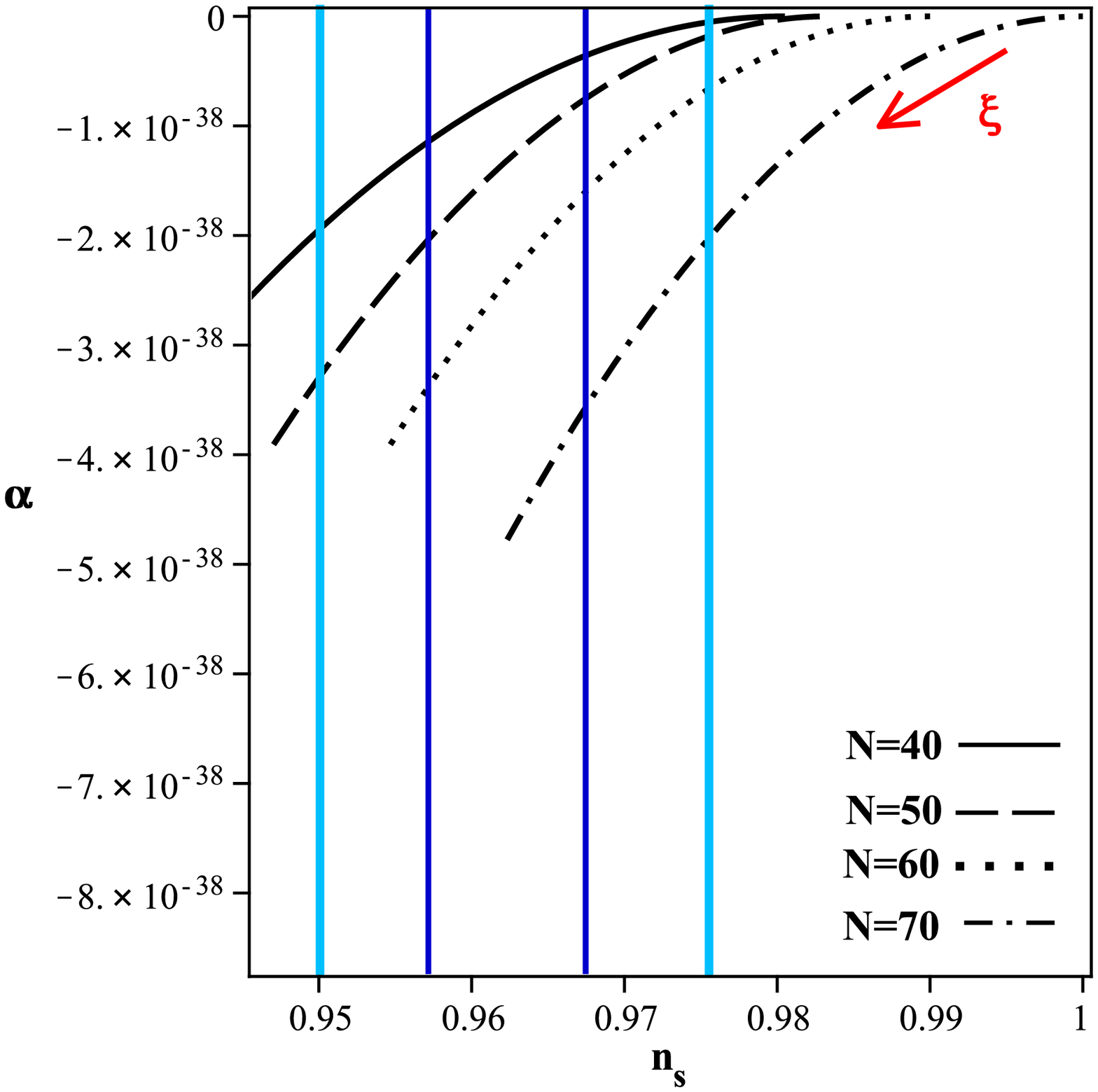} \vspace{6.5cm}
\end{center}
\caption{\label{fig:1}\small{Evolution of the tensor to scalar ratio
(left panel) and running of the scalar spectral index (right panel)
versus the scalar spectral index, for a non-minimally coupled DGP
model with a quadratic potential, in the background of
Planck+WMAP9+BAO data. The figure has been plotted for four values
of the number of e-folds. The non-minimal coupling parameter, $\xi$,
increases in the direction of the red arrow. For $N=70$, the
non-minimally coupled DGP setup is well outside the $95\%$ CL of the
Planck+WMAP9+BAO data for all values of $\xi$. For other values of
$N$, the model lies inside the $95\%$ CL of the Planck+WMAP9+BAO
data for some values of $\xi$.}}
\end{figure}

The first potential which we consider, is a quadratic potential. In
[26] it has been shown that in 4-dimension the model with quadratic
potential lies outside the $95\%$ CL of the joint Planck+WMAP9+BAO
data for $N=50$ and inside it for $N=60$. Now we explore the
situation for a 5-dimension model. At the time of writing our
previous work [14], we had used the observational data of WMAP7.
According to the WMAP7+BAO+H$_{0}$ data [30], a warped DGP model
with minimally coupled scalar field and with a quadratic potential,
lies inside the $95\%$ CL for $N< 70$. For non-minimal coupling case
with $N=70$, the model is within $95\%$ CL of WMAP7+BAO+H$_{0}$
data. Now, with recent Planck date, the situations change
considerably. In a minimally coupled DGP model with a quadratic
potential, for all $N\geq 40$ the model is outside the joint $95\%$
CL of the Planck+WMAP9+BAO data. In a non-minimally coupled DGP
setup, for all $N\geq 70$ the model is well outside the joint $95\%$
CL of the Planck+WMAP9+BAO data. For $N<70$, in some range of $\xi$
the model is compatible with observation. For $N=60$, the model with
$0.051\leq\xi<0.09$, for $N=50$ the model with
$0.032\leq\xi\leq0.0784$ and for $N=40$ the model with
$0.011<\xi<0.0553$ lies inside the $95\%$ CL of the joint
Planck+WMAP9+BAO data. The left panel of figure \ref{fig:1} shows
the behavior of the tensor to scalar ratio versus the scalar
spectral index in the background of the Planck+WMAP9+BAO data. The
figure has been plotted for four values of $N$ and for $\xi\geq 0$.
We also, have plotted the evolution of the running of the spectral
index versus the scalar spectral index in the background of the
Planck+WMAP9+BAO data (the left panel of figure \ref{fig:1}). We see
that, for all four values of the number of e-folds, the running of
the scalar spectral index is negative and very close too zero. Note
that in plotting $r$ versus $n_{s}$ we take $k=0.002Mpc^{-1}$ and in
plotting $\alpha$ versus $n_{s}$ we set $k=0.038$, according to
[26].

\subsection{$V(\varphi)=\frac{\sigma}{4}\varphi^{4}$}

\begin{figure}[htp]
\begin{center}\includegraphics{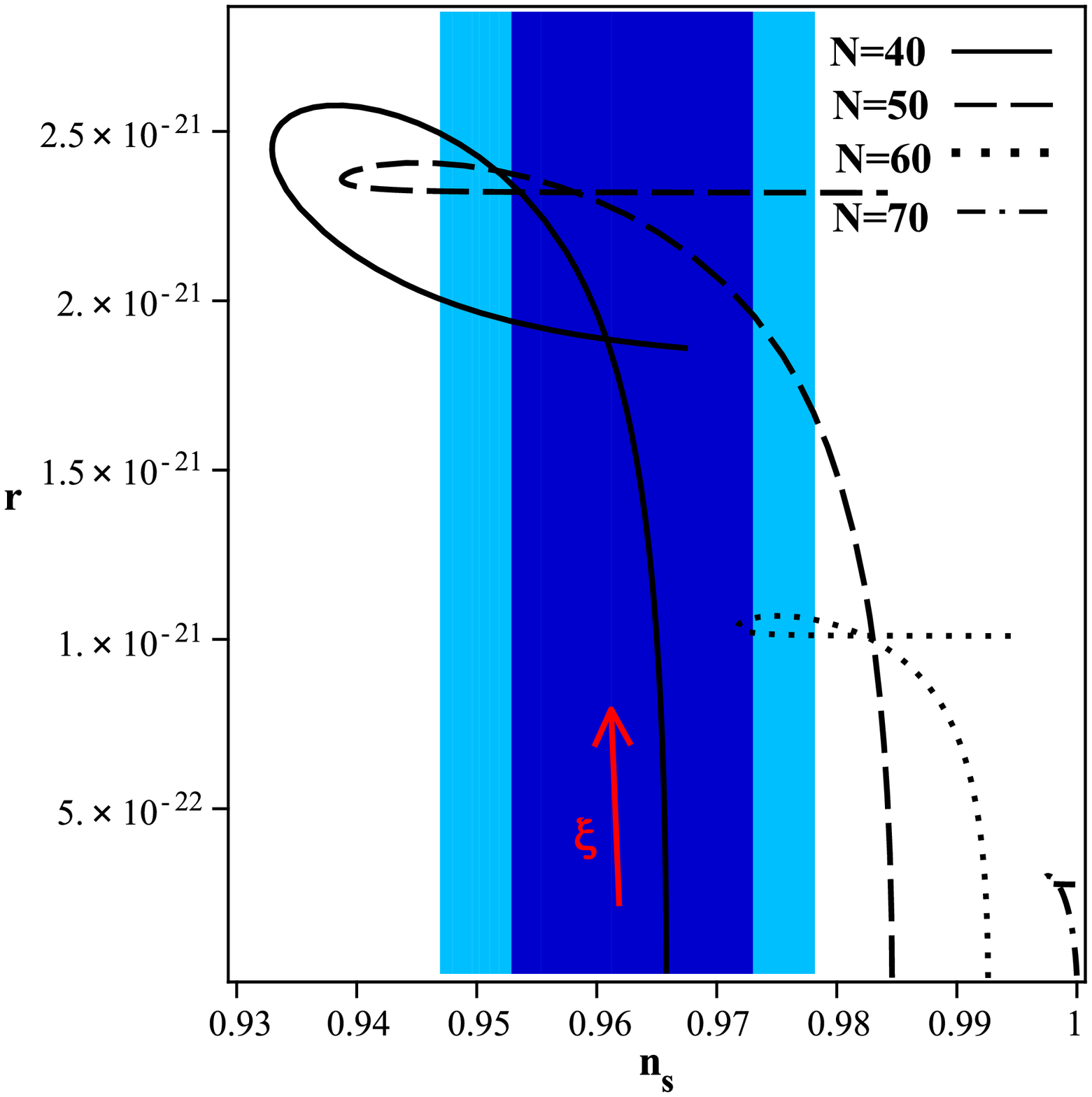} \includegraphics{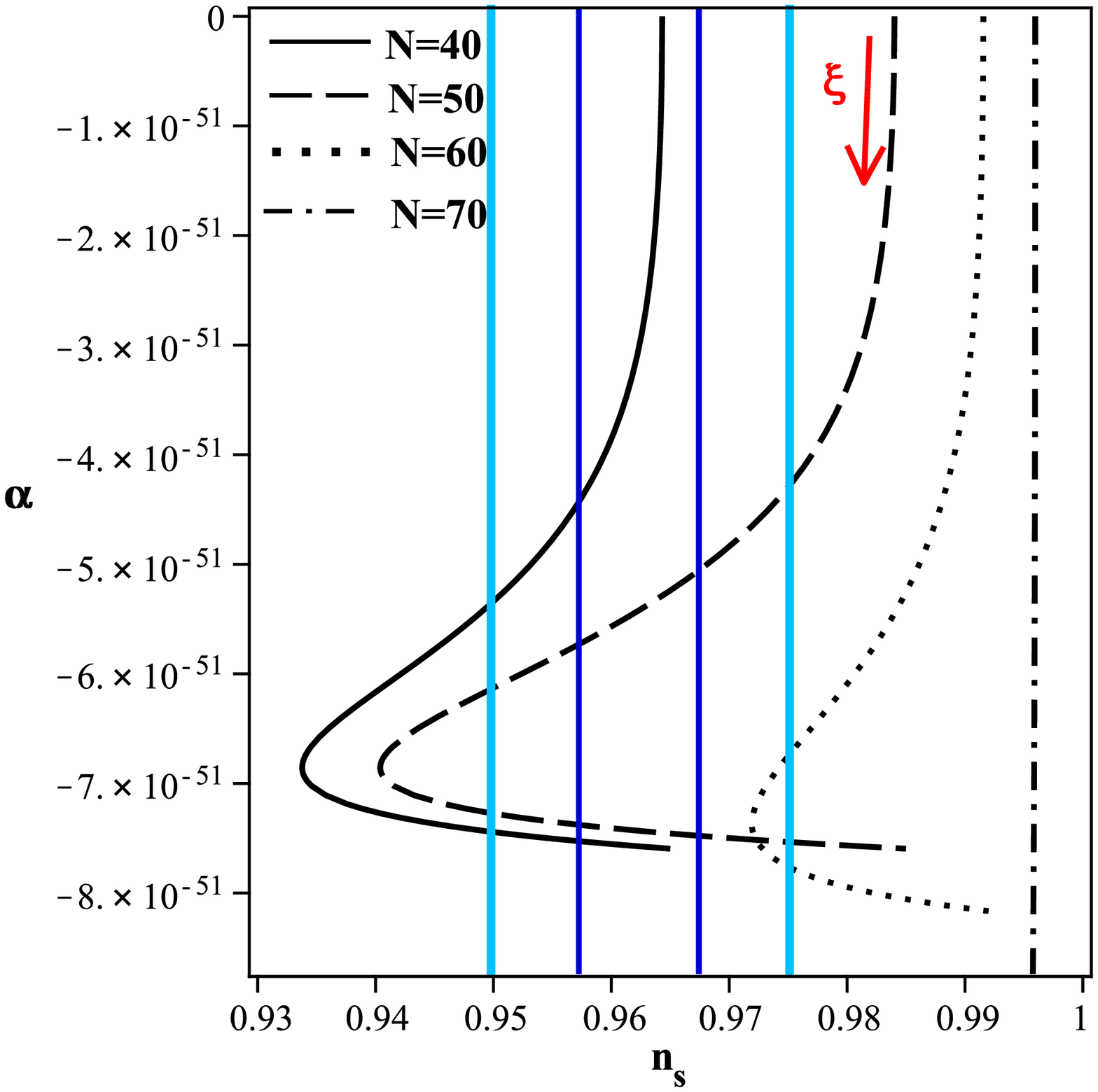} \vspace{6.5cm}
\end{center}
\caption{\label{fig:2}\small{Evolution of the tensor to scalar ratio
(left panel) and running of the scalar spectral index (right panel)
versus the scalar spectral index, for a non-minimally coupled DGP
model with a quartic potential, in the background of
Planck+WMAP9+BAO data. The figure has been plotted for
$N=40,\,50,\,60\,$ and $70$ and with $\xi\geq0$.  For $N=70$, the
non-minimally coupled DGP setup is well outside the $95\%$ CL of the
Planck+WMAP9+BAO data for all values of $\xi$. For other values of
$N$, the model lies inside the $95\%$ CL of the Planck+WMAP9+BAO
data in some range of $\xi$.}}
\end{figure}

The next potential we consider, is a quartic potential. It has been
confirmed with WMAP9 [31] and Planck [26] data that, a model with a
quartic potentia in 4-dimension lies outside the $95\%$ CL. In our
branworld setup we reach a different result: a minimally coupled DGP
model with this potential, lies inside the $95\%$ CL of
WMAP7+BAO+H$_{0}$ data for $30\leq N\leq60$, and lies inside the
$95\%$ CL of the Planck+ WMAP9+ BAO data for $30\leq N\leq46$. If we
consider a non-minimally coupled scalar field on the DGP brane, the
model lies inside the $95\%$ CL of the Planck+WMAP9+BAO data for
$N>46$ too. We have studied the behavior of the inflationary
parameters in the non-minimally coupled DGP model for
$N=40,\,50,\,60\,$ and 70 and for $\xi\geq0$ in the background of
Planck+WMAP9+BAO data. The result is shown in figure \ref{fig:2}.
For $N=70$, with all values of $\xi$, the model is well outside the
$95\%$ CL of the Planck+WMAP9+BAO data. For $N=60$, the model with
$0.09\leq\xi\leq0.108$ lies inside the $95\%$ CL of the data. For
$N=50$, the model with $0.51\leq\xi\leq0.0891$ and
$0.11<\xi\leq0.131$ is compatible with the Planck+WMAP9+BAO data.
For $N=40$, the model with $\xi\leq0.101$ and $\xi\geq0.1308$ is
well inside the the $95\%$ CL of the Planck+WMAP9+BAO data. Note
that the evolution of the running of the scalar spectral index
corresponding to the quartic potential is shown in the right panel
of figure \ref{fig:2}. The value of the running of the spectral
index is negative and very close to zero for all values of $\xi$.

\begin{figure}[htp]
\begin{center}\includegraphics{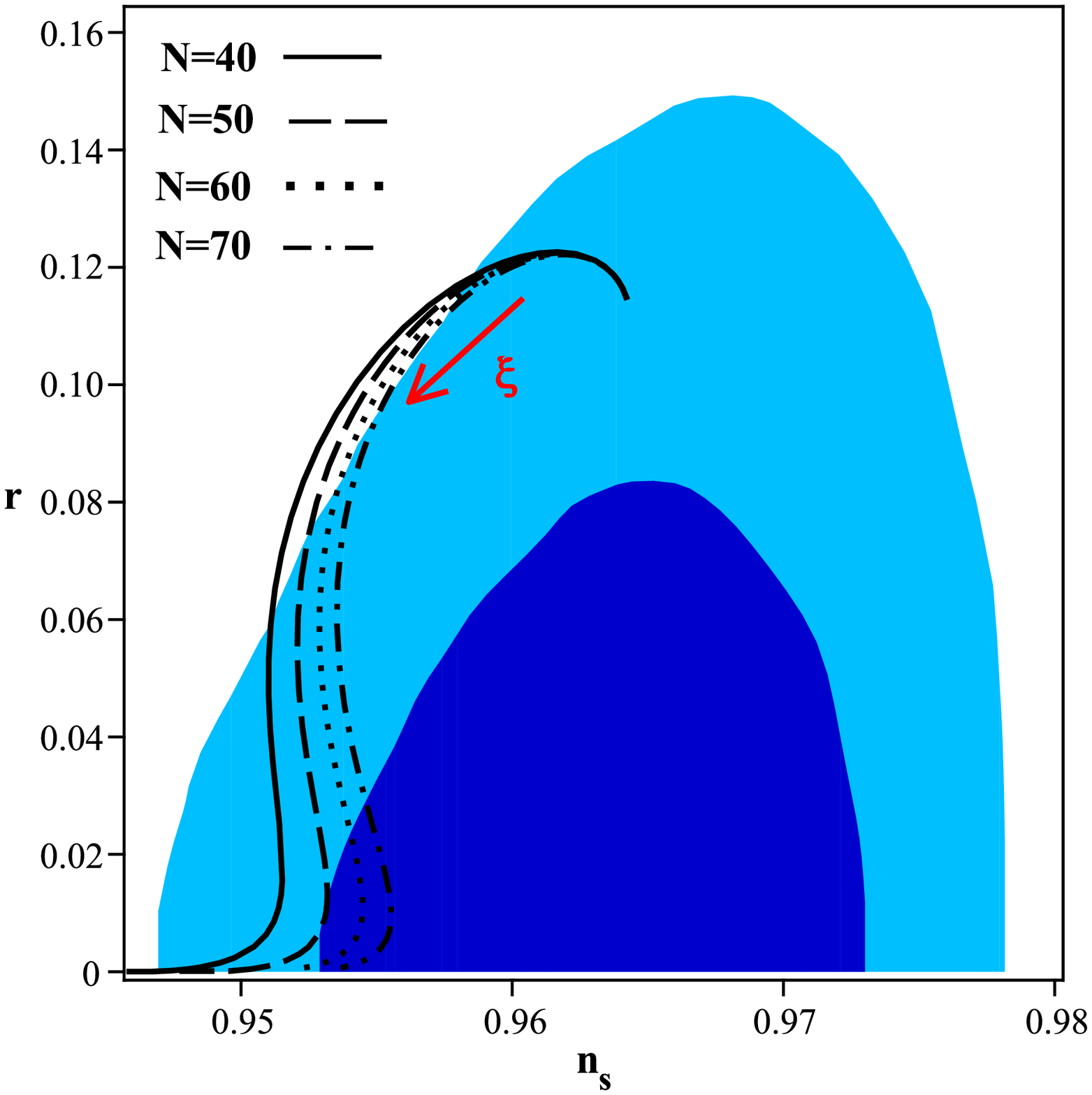} \includegraphics{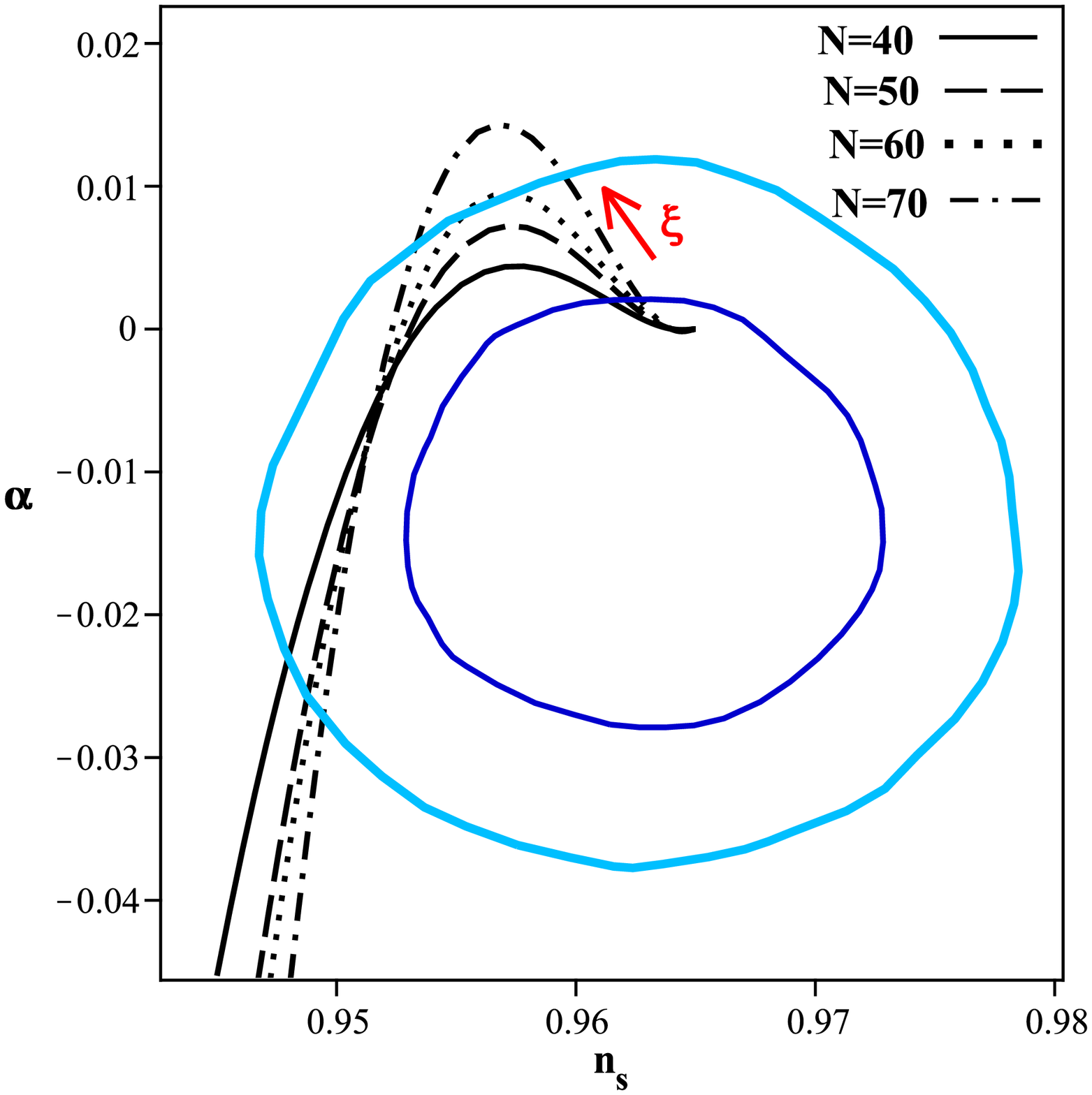} \vspace{6.5cm}
\end{center}
\caption{\label{fig:3}\small{Evolution of the tensor to scalar ratio
(left panel) and running of the scalar spectral index (right panel)
versus the scalar spectral index, for a non-minimally coupled DGP
model with a linear potential, in the background of Planck+WMAP9+BAO
data. The figure has been plotted for $N=40,\,50,\,60\,$ and $70$
and with $\xi\geq0$. For all given values of the number of e-folds,
the non-minimally coupled DGP setup in some range of $\xi$ lies
inside the $95\%$ CL of the Planck+WMAP9+BAO.}}
\end{figure}

\subsection{$V(\varphi)=\sigma\varphi$}
Now, we consider a linear inflationary potential as
$V=\sigma\varphi$ [32] which is motivated by axion monodromy. A
minimally coupled four-dimensional setup with this potential lies
within the $95\%$ CL of the Planck+WMAP9+BAO data [26]. Our
braneworld model (either minimally or non-minimally coupled setup),
with this linear potential, lies within the $95\%$ CL of
WMAP7+BAO+H$_{0}$. A minimally coupled DGP model with this potential
lies still inside the $95\%$ CL of the Planck+WMAP9+BAO data. But,
there are constraints on the parameters in the non-minimal coupling
case. As other cases, we consider four values of number of e-folds.
In the left panel of figure \ref{fig:3}, we see the evolution of the
tensor to scalar ratio versus the scalar spectral index with
$\xi\geq0$. Our numerical analysis shows that, a DGP model with a
non-minimally coupled scalar field, in some range of $\xi$ is
observationally viable. For $N=70$, the model with $\xi\leq0.021$
and $0.3201\leq\xi$ lies inside the $95\%$ CL of the
Planck+WMAP9+BAO data. For $N=60$, the model with $\xi\leq0.0195$
and $0.0383\leq\xi$ is compatible with data. For $N=50$, the
non-minimally coupled DGP model with $\xi\leq0.0193$ and
$0.0461\leq\xi$ is within the $95\%$ CL of the Planck+WMAP9+BAO
data. Finally, for $N=40$, the model with $\xi<0.0192$ and
$0.0602<\xi<0.1301$ lies within the range of observational data. The
right panel of figure \ref{fig:3} shows the evolution of the running
of the scalar spectral index versus the scalar spectral index. As
the figure shows, for a non-minimally coupled DGP model with a
linear potential, the running of the scalar spectral index is
negative. The value of running in this case is relatively large.

\begin{figure}[htp]
\begin{center}\includegraphics{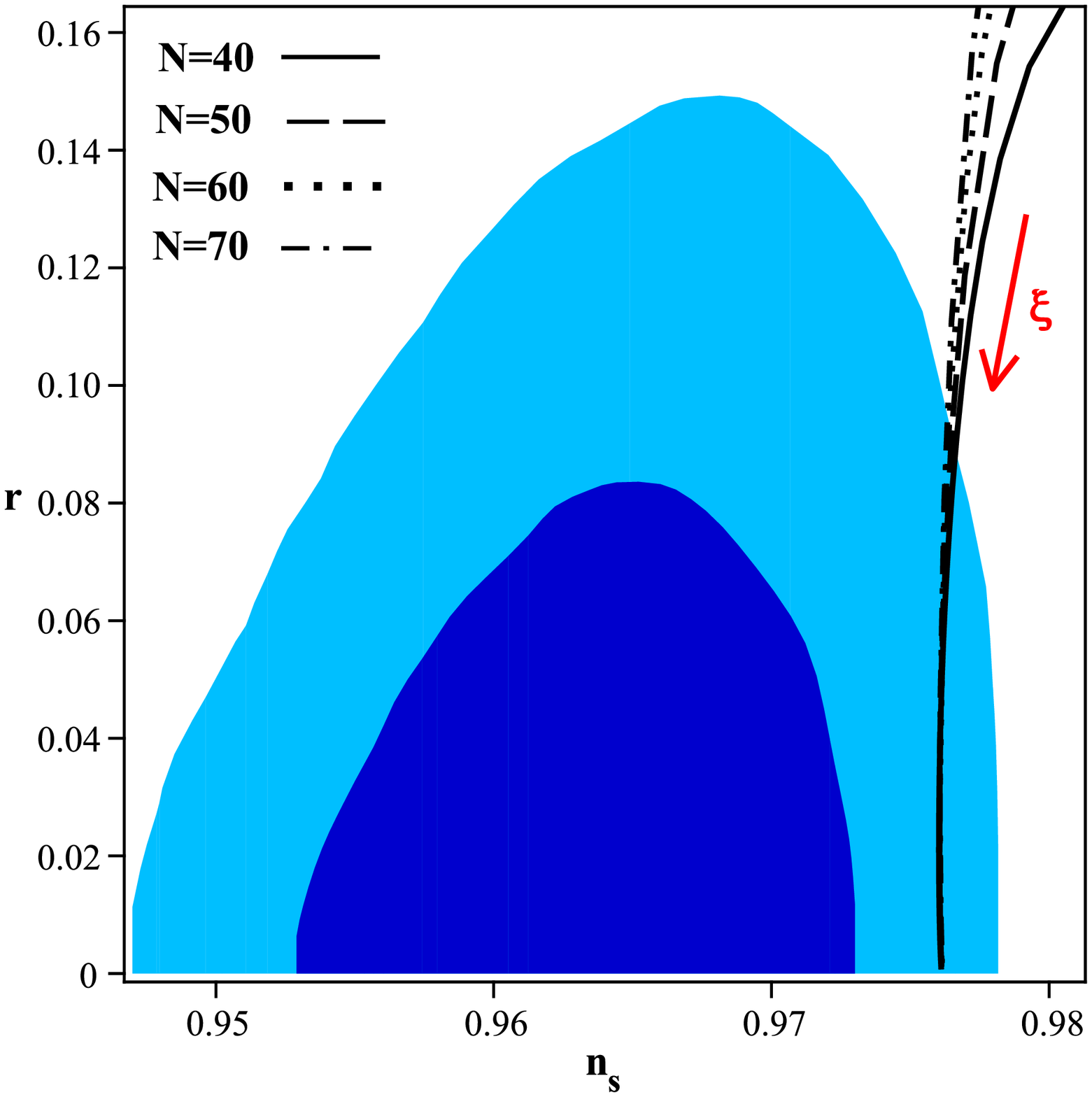} \includegraphics{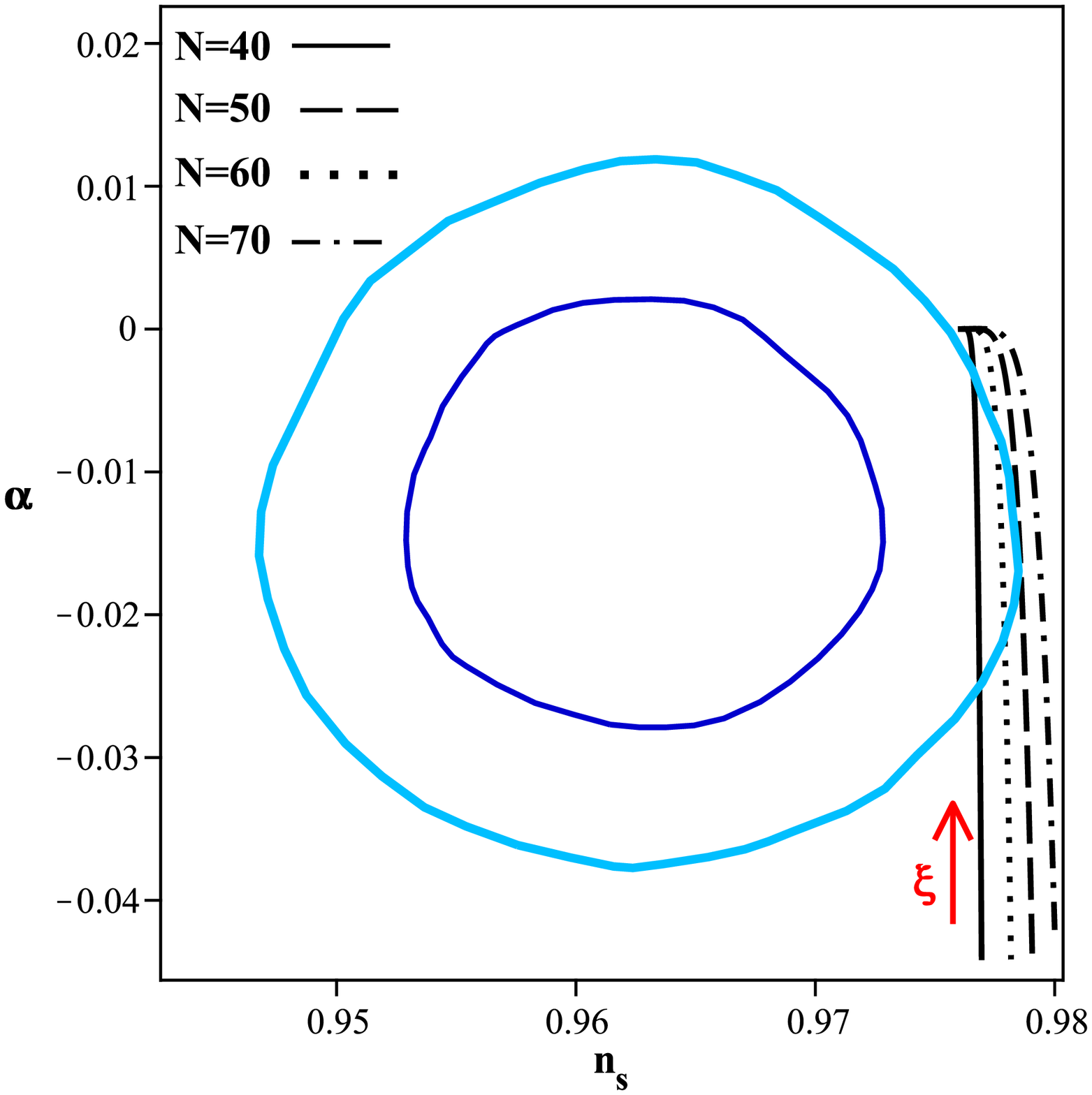} \vspace{6.5cm}
\end{center}
\caption{\label{fig:4}\small{Evolution of the tensor to scalar ratio
(left panel) and running of the scalar spectral index (right panel)
versus the scalar spectral index, for a non-minimally coupled DGP
model with a potential of type
$V(\varphi)\sim\varphi^{\frac{2}{3}}$, in the background of
Planck+WMAP9+BAO data. The figure has been plotted for
$N=40,\,50,\,60\,$ and $70$ and with $\xi\geq0$. For all given
values of the number of e-folds, the non-minimally coupled DGP setup
in some range of $\xi$ lies inside the $95\%$ CL of the
Planck+WMAP9+BAO  data.}}
\end{figure}

\subsection{$V(\varphi)=\sigma\varphi^{\frac{2}{3}}$}
In this subsection we adopt another type of potential motivated by
axion monodromy, as $V(\varphi)=\sigma\varphi^{\frac{2}{3}}$ [33].
As shown in Ref. [26], a minimally coupled 4-dimensional model with
this type of potential lies within the $95\%$ CL of the
Planck+WMAP9+BAO data. Although a minimally coupled DGP model with
this potential lies in the $95\%$ CL of the WMAP7+BAO+H$_{0}$, it is
now well outside the $95\%$ CL of the Planck+WMAP9+BAO data. If we
consider a non-minimally coupled scalar field on the DGP brane, we
find that the model in some range of $\xi$ is compatible with newly
released observational data. The evolution of the tensor to scalar
ratio versus the scalar spectral index is shown in the left panel of
figure \ref{fig:4}. For $N=70$, the non-minimally coupled DGP model
with $\xi\geq0.076$, for $N=60$, the model with $\xi\geq0.0768$, for
$N=50$, the model with $\xi\geq0.0775$ and for $N=40$, the model
with $\xi\geq0.078$ lies inside the $95\%$ CL of the
Planck+WMAP9+BAO data. Note that, as $\xi$ increases the
non-minimally coupled DGP model with this potential becomes
independent of the value of $N$. We have also plotted the evolution
of the running of the scalar spectral index versus the scalar
spectral index (right panel of figure \ref{fig:4}). For this case,
the running is negative and in some range of $\xi$ is compatible
with recent data.

\begin{figure}[htp]
\begin{center}\includegraphics{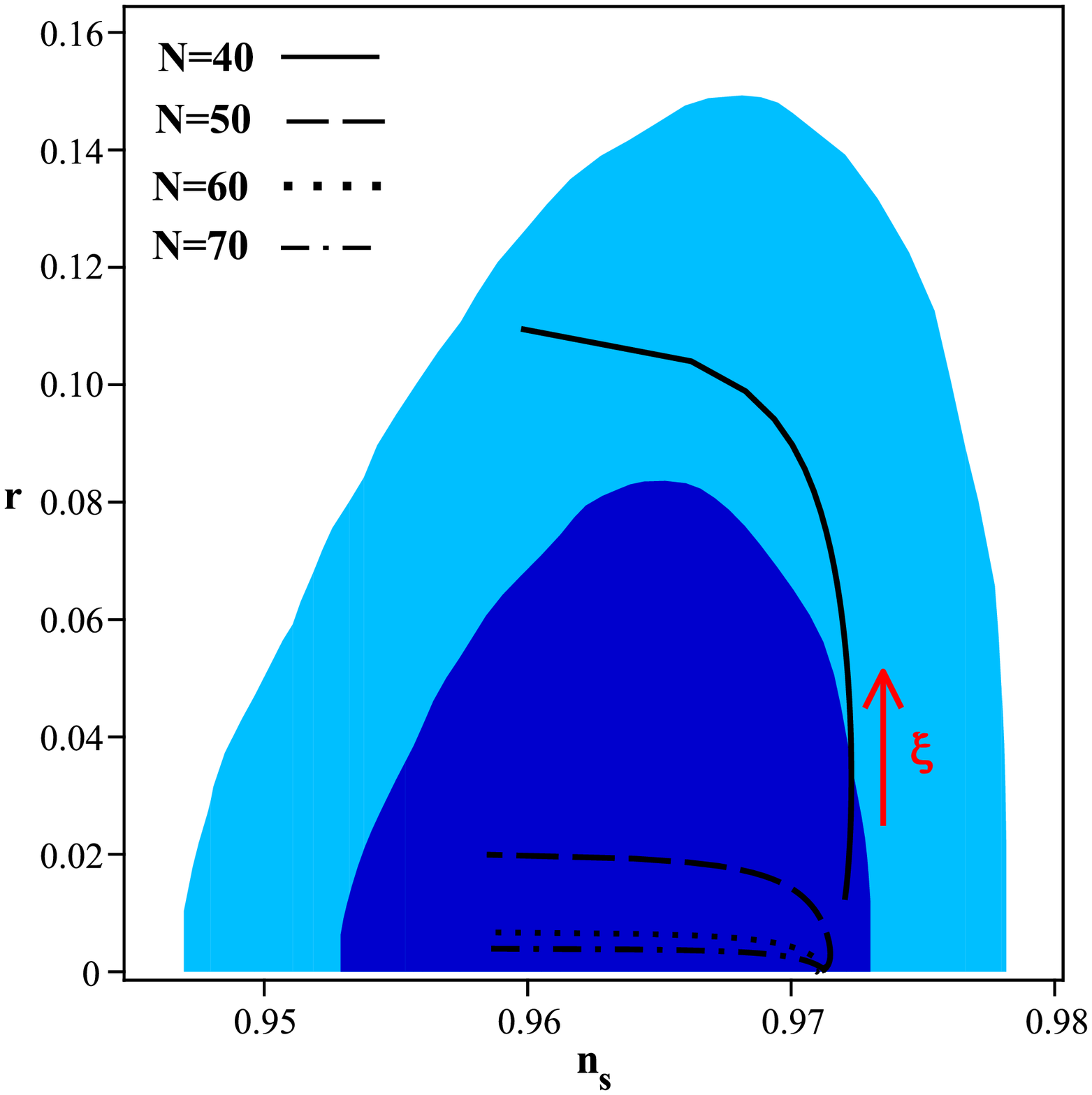} \includegraphics{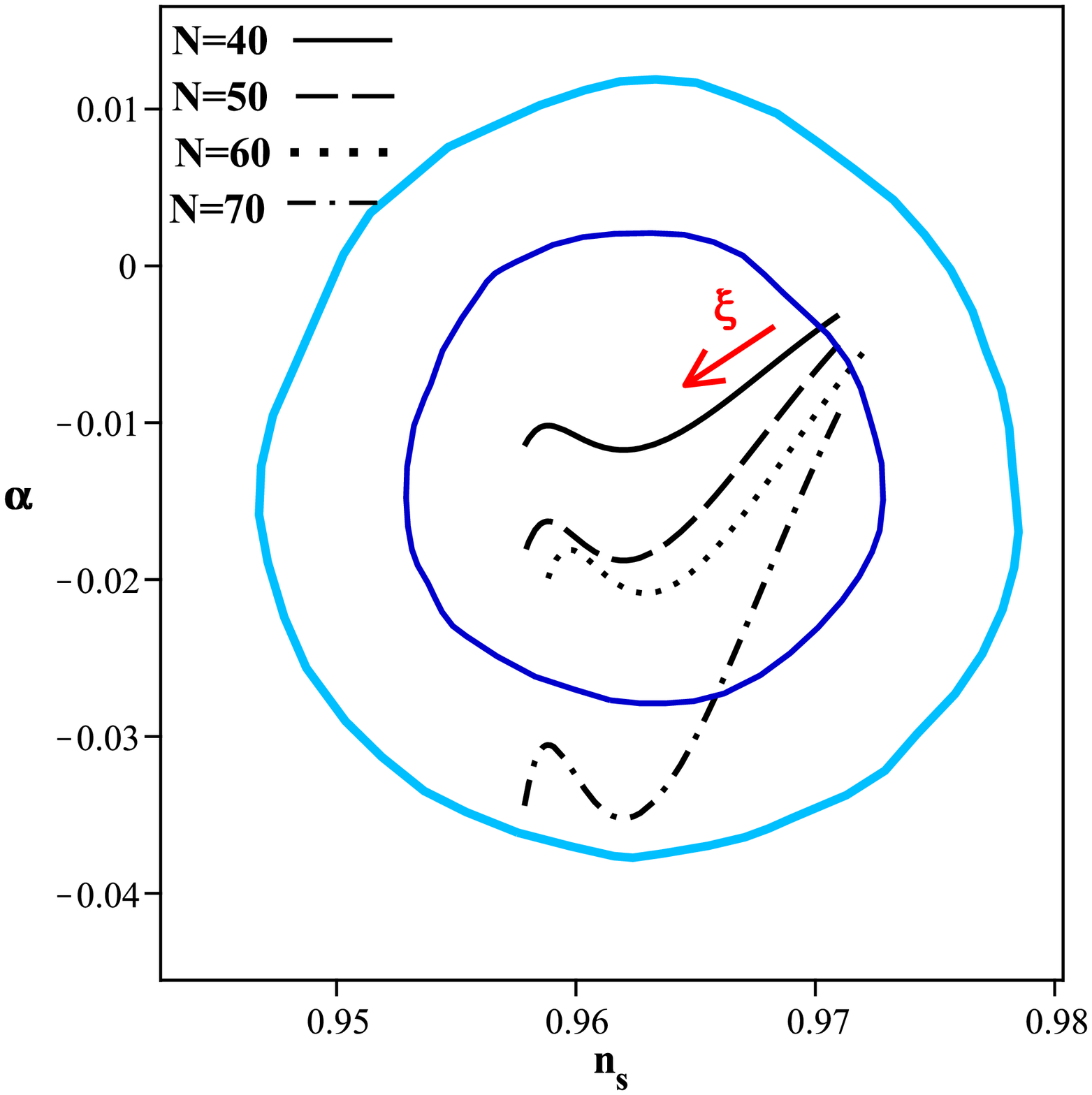} \vspace{6.5cm}
\end{center}
\caption{\label{fig:5}\small{Evolution of the tensor to scalar ratio
(left panel) and running of the scalar spectral index (right panel)
versus the scalar spectral index, for a non-minimally coupled DGP
model with an exponential potential of the type
$V(\varphi)\sim\exp(-\beta\varphi)$, in the background of the
Planck+WMAP9+BAO data. The figure has been plotted for
$N=40,\,50,\,60\,$ and $70$ and with $\xi\geq0$. For all given
values of the number of e-folds, the non-minimally coupled DGP setup
with all values of $\xi$ lies inside the $95\%$ CL of the
Planck+WMAP9+BAO data.}}
\end{figure}

\subsection{$V(\varphi)=V_{0}\exp(-\beta\varphi)$}
Here, we study the inflationary parameter with an exponential
potential where $\beta$ is a positive constant. With this type of
potential, a 4-dimensional model is well outside the $95\%$ CL
(even, $0.99.7\%$ CL) of the Planck+WMAP9+BAO data (see [26]). For a
DGP model with this potential and with either minimally or
non-minimally coupled scalar field on the brane, we reach different
results. This model with an exponential potential with all values of
$\xi$ lies inside the $95\%$ CL of the Planck+WMAP9+BAO data. In the
left panel of figure \ref{fig:5}, we see the evolution of the tensor
to scalar ratio versus the scalar spectral index in the background
of the Planck+WMAP9+BAO data. As we see the models with number of
e-folds $N=50,\,60,\, 70$ give more reasonable results than the
$N=40$ case. The right panel of figure \ref{fig:5} shows the
behavior of the running of the spectral index versus the spectral
index. With an exponential potential, the running is large and
negative.

\subsection{$V(\varphi)=V_{0}\exp(\beta\varphi)$}
\begin{figure}[htp]
\begin{center}\includegraphics{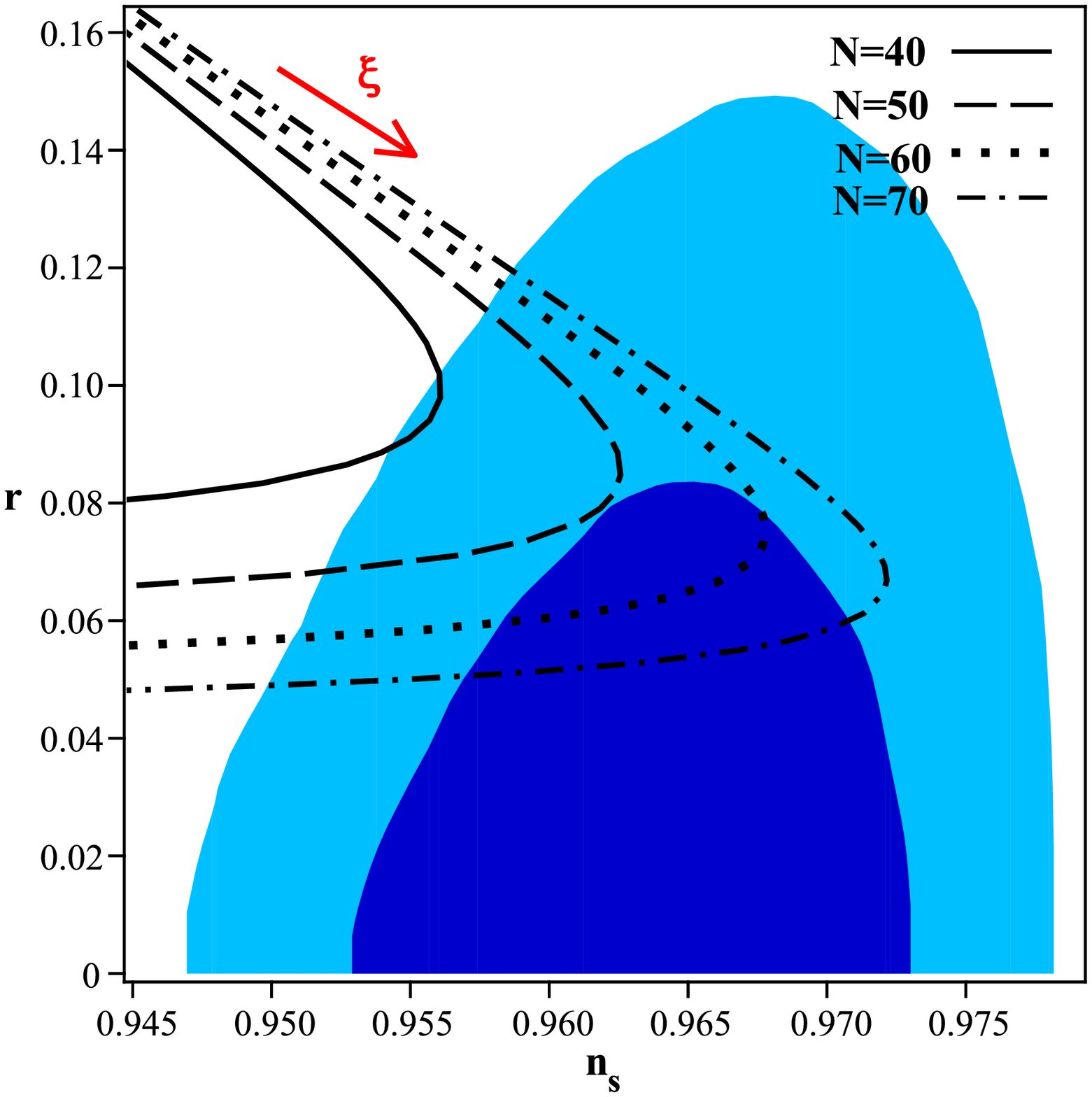} \includegraphics{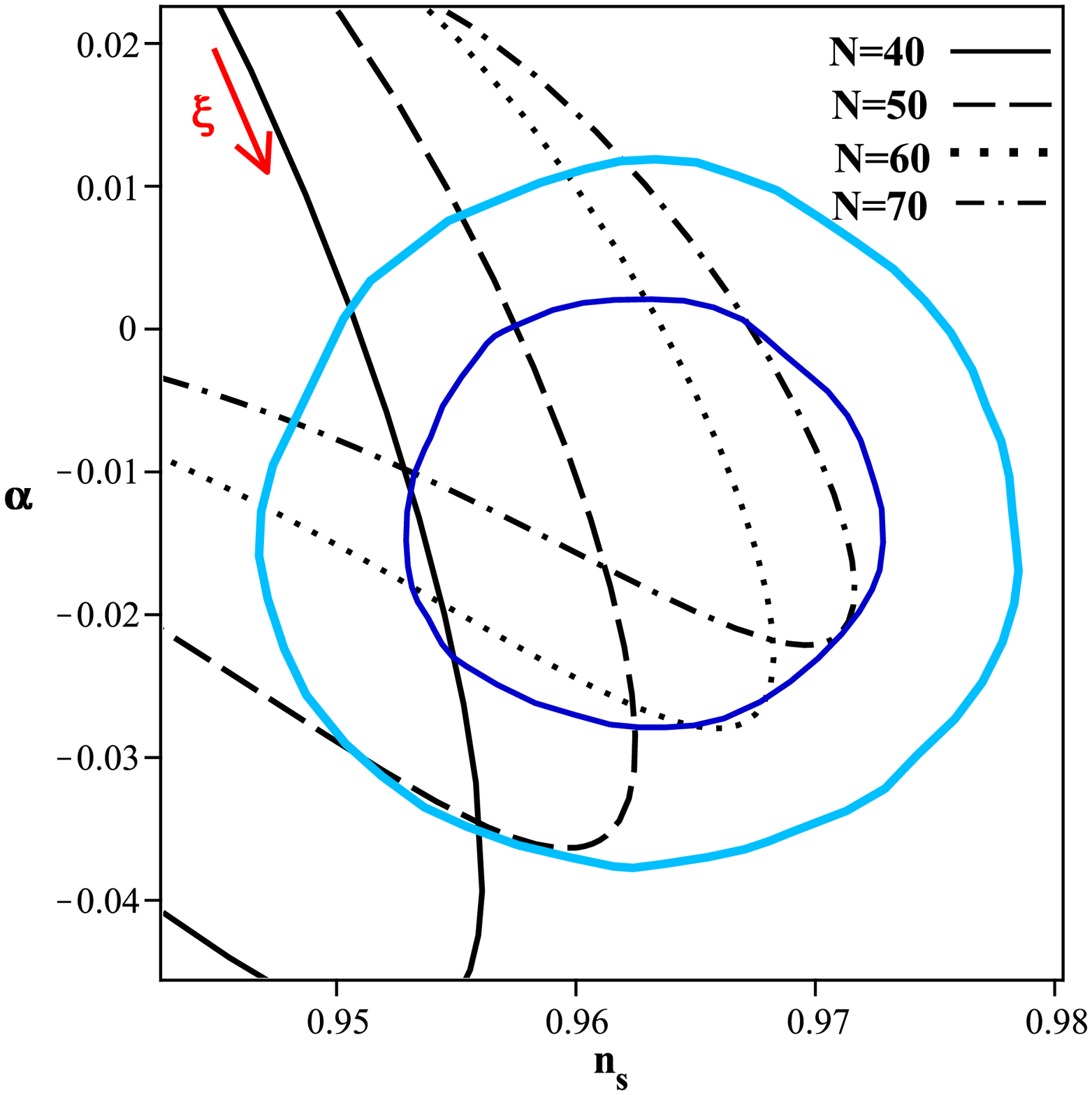} \vspace{6.5cm}
\end{center}
\caption{\label{fig:6}\small{Evolution of the tensor to scalar ratio
(left panel) and running of the scalar spectral index (right panel)
versus the scalar spectral index, for a non-minimally coupled DGP
model with an exponential potential of the type
$V(\varphi)\sim\exp(\beta\varphi)$, in the background of the
Planck+WMAP9+BAO data. The figure has been plotted for
$N=40,\,50,\,60\,$ and $70$ and with $\xi\geq0$. For all given
values of the number of e-folds, the non-minimally coupled DGP setup
in some range of $\xi$ lies inside the $95\%$ CL of the
Planck+WMAP9+BAO data. Also, the running of the scalar spectral
index is negative just in some ranges of $\xi$.}}
\end{figure}

In this subsection we consider another type of exponential potential
with positive exponent. A minimally coupled DGP model with this
potential lies outside the the $95\%$ CL of the WMAP7+BAO+H$_{0}$
data and even $95\%$ CL of the Planck+WMAP9+BAO data. However, a
non-minimally coupled DGP model with this exponential potential, in
some range of $\xi$, is compatible with observational data. The left
panel of the figure \ref{fig:6} shows the behavior of the tensor to
scalar ratio versus the scalar spectral index with $\xi\geq0$. For,
$N=70$, the non-minimally coupled DGP model with
$0.0431\leq\xi\leq0.1301$, for $N=60$, the model with
$0.0442<\xi\leq0.127$, for $N=50$, the model with
$0.052\leq\xi\leq0.101$ and for $N=40$, the model with
$0.058\leq\xi\leq0.0621$ lies inside the $95\%$ CL of the
Planck+WMAP9+BAO data. The left panel of the figure \ref{fig:6}
shows the evolution of the running of the spectral index versus the
scalar spectral index. With this exponential potential, the running
is negative in some range of $\xi$ and is positive in other ranges.
For $N=70$ the running of the spectral index with $\xi\geq0.046$,
for $N=60$ the running with $\xi\geq0.04201$, for $N=50$ the running
with $\xi\geq0.0372$ and for $N=40$ the running with $\xi\geq0.0291$
is negative.

In table \ref{tab:1}, we have summarized the result of our numerical
analysis in a non-minimally coupled DGP model. We have clearly
specified the range of the non-minimal coupling parameter in which
the model is compatible with Planck+WMAP9+BAO data.

\begin{table*}
\caption{\label{tab:1}The range of $\xi$ for which a non-minimally
coupled DGP model lies inside the Planck+WMAP9+BAO data.}
\begin{tabular}{ccccc}
\\ \hline \hline$V$&$N=40$&$N=50$&$N=60$&$N=70$ \\ \hline\\
$\frac{\sigma}{2}\varphi^{2}$& $0.011<\xi<0.0553$ &$0.032\leq\xi\leq 0.0784$&$0.051\leq\xi<0.09$&---\\\\
$\frac{\sigma}{4}\varphi^{4}$& $\xi\leq0.101$ &$0.51\leq \xi\leq
0.0891$& $0.09 \leq\xi \leq 0.108$& ---\\
&and $\xi\geq0.1308$ &and $0.11< \xi \leq0.131$
& & \\\\
$\sigma\varphi$& $\xi < 0.0192$ & $\xi \leq 0.0193$
& $\xi\leq0.0195$& $\xi\leq 0.021$\\
&and $0.0602 <\xi<0.1301$ &and $0.0461\leq \xi$
& and $0.0383 \leq \xi$& and $0.3201 \leq \xi$\\\\
$\sigma\varphi^{\frac{2}{3}}$& $0.078\leq\xi$ &$0.078\leq\xi$& $0.0768\leq\xi$ & $0.076\leq\xi$\\\\
$V_{0}e^{-\beta\varphi}$& all values of $\xi$ & all values of $\xi$ & all values of $\xi$ & all values of $\xi$\\\\
$V_{0}e^{\beta\varphi}$& $0.058 \leq\xi \leq 0.0621$ &$0.052 \leq\xi
\leq 0.101$&
$0.0442 < \xi \leq 0.127$& $0.0431 \leq\xi \leq 0.1301$\\ \hline\\\\
\end{tabular}
\end{table*}

\section{Conclusion}
In this paper we have studied the observational status of a
non-minimal inflation model on a warped DGP braneworld scenario. By
choosing some well-motivated types of potential, we have studied the
evolution of the inflationary parameters in the background of the
newly released Planck+WMAP9+BAO data. We have also, compared our
result with those results obtained in a 4-dimensional setup and we
found that the presence of the non-minimal coupling and the extra
dimension causes considerable differences relative to the four
dimensional case. Although a minimal 4D model with a quadratic
potential lies inside the $95\%$ CL of the Planck+WMAP9+BAO data, a
minimal DGP model with this potential is outside the data range.
But, if we include an explicit non-minimal coupling between the
scalar field and induced gravity on the brane, we find that the
model with specific values of $N$ and in some ranges of the
non-minimal coupling parameter lies well in the $95\%$ CL of the
Planck+WMAP9+BAO data. With a quartic potential, the 4-dimensional
model is well outside the $95\%$ CL of the Plank+WMAP9+BAO data
whereas, the DGP model with specific values of $N$ lies within the
$95\%$ CL of the Planck+WMAP9+BAO data. A non-minimally coupled DGP
model in larger range of $N$ is compatible with Planck data. Both
minimal 4D model and minimal DGP model, with a linear potential, lie
inside the Planck data. But, a non-minimal DGP model with a linear
potential is outside the $95\%$ CL of the Planck+WMAP9+BAO data, in
some range of $\xi$. A minimal four dimensional model with potential
of the type $V\sim\varphi^{\frac{2}{3}}$ lies inside the $95\%$ CL
of the Planck+WMAP9+BAO data whereas, a minimal DGP model with this
type of potential is outside the range of these observational data.
In the presence of the non-minimal coupling, the DGP model in some
range of $\xi$ lies inside the $95\%$ CL of the Planck+WMAP9+BAO
data. It seems that, an exponential potential
($V\sim\exp(-\beta\varphi)$) is the best potential for a DGP model
in an inflationary stage. With this potential, both minimally and
non-minimally coupled DGP setup lie well inside the the $95\%$ CL of
the Planck+WMAP9+BAO data. Note that, a four-dimensional model with
this potential is quite outside the Planck data [26]. The last
potential which we have considered in this paper is an exponential
potential as $V\sim\exp(\beta\varphi)$. With this potential, a
minimal DGP model lies outside the Planck data and a non-minimal DGP
model, in some range of $\xi$ lies inside the $95\%$ CL of the
Planck+WMAP9+BAO data. In summary, within a braneworld setup with
induced gravity and a non-minimally coupled scalar field on the
brane, a inflation model with $V\sim\exp(-\beta\varphi)$ has the
best fit with the recent observational data.


\begin{thebibliography}{11}

\bibitem{1}
A. Guth, Phys. Rev. D, \textbf{62}, 105030, (1981).

\bibitem{2}
A. D. Linde, Phys. Lett. B, \textbf{108}, 389 (1982)

\bibitem{3}
A. Albrecht and P. Steinhard, Phys. Rev. D, \textbf{48}, 1220,
(1982)

\bibitem{4}
A. D. Linde, \emph{Particle Physics and Inflationary Cosmology}
(Harwood Academic Publishers, Chur, Switzerland, 1990).

\bibitem{5}
A. Liddle and D. Lyth, \emph{Cosmological Inflation and Large-Scale
Structure}, (Cambridge University Press, 2000).

\bibitem{6}
J. E. Lidsey et al, Phys. Rev. D, \textbf{69}, 373, (1997).

\bibitem{7}
A. Riotto, [arXiv:hep-ph/0210162].

\bibitem{8}
D. H. Lyth and A. R. Liddle, \emph{The Primordial Density
Perturbation} (Cambridge University Press, 2009).

\bibitem{9}
R. H. Brandenberger, [arXiv:hep-th/0509099].

\bibitem{10}
R. Maartens, D. Wands, B. Bassett, I. Heard, Phys. Rev. D,
\textbf{62}, 041301, (2000).

\bibitem{11}
R. Cai and H. Zhang, JCAP, \textbf{0408}, 017, (2004)

\bibitem{12}
S. del Campo and R. Herrera, Phys. Lett. B, 653, \textbf{122},
(2007).

\bibitem{13}
K. Nozari, M. Shoukrani and B. Fazlpour, Gen. Rel. Grav,
\textbf{43}, 207, (2011).

\bibitem{14}
K. Nozari and N. Rashidi, Phys. Rev. D, \textbf{86}, 043505, (2012).

\bibitem{15}
V. Faraoni, Phys. Rev. D, \textbf{53}, 6813, (1996).

\bibitem{16}
V. Faraoni, Phys. Rev. D, \textbf{62}, 023504, (2000).

\bibitem{17}
V. Faraoni, Int. J. Theor. Phys., \textbf{38}, 217, (1999).

\bibitem{18}
R. Fakir, S. Habib and W. G. Unruh, Astrophys. J., \textbf{394},
396, (1992).

\bibitem{19}
M. V. Libanov, V. A. Rubakov and P. G. Tinyakov, Phys. Lett. B,
\textbf{442}, 63, (1998).

\bibitem{20}
J. Hwang and H. Noh, Phys. Rev. D, \textbf{60}, 123001, (1999).

\bibitem{21}
S. Tsujikawa, K. Maeda and T. Torii, Phys. Rev. D, \textbf{60},
063515, (1999a).

\bibitem{22}
K. Nozari and S. Shafizadeh, Phys. Scr. \textbf{82}, 015901, (2010).

\bibitem{23}
S. Nojiri, S. D. Odintsov [arXiv:0807.0685].

\bibitem{24}
G. Cognola, E. Elizalde, S. Nojiri, S.D. Odintsov, L. Sebastiani and
S. Zerbini, Phys. Rev. D \textbf{77}, 046009, (2008).

\bibitem{25}
S. Nojiri, S. D. Odintsov, Phys. Rev. D, \textbf{77}, 026007,
(2008).

\bibitem{26}
P. A. R. Ade {\it et al.}, [arXiv:1303.5082].

\bibitem{27}
J. Bardeen,  Phys. Rev. D, \textbf{22}, 1882, (1980).

\bibitem{28}
V. F. Mukhanov, H. A. Feldman and R. H. Brandenberger, Phys. Rept.,
\textbf{215}, 203, (1992).

\bibitem{29}
E. Bertschinger, [arXiv:astro-ph/9503125].

\bibitem{30}
E. Komatsu {\it et al.}, Astrophys. J. Suppl., \textbf{192}, 18,
(2011).

\bibitem{31}
Hinshaw  {\it et al.}, [arXiv:1212.5226], (2013).

\bibitem{32}
L. McAllister, E. Silverstein and A. Westphal, Phys. Rev., D
\textbf{82}, 046003, (2010).

\bibitem{33}
E. Silverstein and A. Westphal, Phys. Rev., D \textbf{78}, 106003,
(2008)

\end{thebibliography}
\end{document}